\documentclass[twocolumn]{aastex631}

\received{XXX}
\revised{YYY}
\accepted{ZZZ}

\shorttitle{Long-term NIR Variability of OJ 287}
\shortauthors{Gupta et al.}

\begin{document}

\title{Long Term Multi-band Near Infra-Red Variability of the Blazar OJ 287 during 2007--2021}

\correspondingauthor{Alok C. Gupta}
\email{acgupta30@gmail.com} 

\author[0000-0002-9331-4388]{Alok C.\ Gupta}
\affiliation{Aryabhatta Research Institute of Observational Sciences (ARIES), Manora Peak, Nainital -- 263001, India}
\affiliation{Shanghai Frontiers Science Center of Gravitational Wave Detection, 800 Dongchuan Road, Minhang, Shanghai 200240, 
People's Republic of China}

\author[0000-0001-6890-2236]{Pankaj Kushwaha}\thanks{Email: pankaj.kushwaha@iisermohali.ac.in}
\altaffiliation{DST-INSPIRE Faculty Fellow}
\affiliation{Department of Physical Sciences, Indian Institute of Science Education and Research Mohali, Knowledge City, Sector 81, SAS Nagar, Punjab 140306, India} 
\affiliation{Aryabhatta Research Institute of Observational Sciences (ARIES), Manora Peak, Nainital -- 263001, India} 

\author{L.\ Carrasco}
\affiliation{Instituto Nacional de Astrof\'isica, \'Optica y Electr\'onica, Luis Enrique Erro 1, Tonantzintla, Puebla, C.P. 72840, M\'exico}

\author{Haiguang Xu}\thanks{E-mail: hgxu@sjtu.edu.cn}
\affiliation{Shanghai Frontiers Science Center of Gravitational Wave Detection, 800 Dongchuan Road, Minhang, Shanghai 200240, 
People's Republic of China}
\affiliation{School of Physics and Astronomy, Shanghai Jiao Tong University, 800 Dongchuan Road, Minhang, Shanghai 200240, 
People's Republic of China} 

\author[0000-0002-1029-3746]{Paul J. Wiita}
\affiliation{Department of Physics, The College of New Jersey, 2000 Pennington Rd., Ewing, NJ 08628-0718, USA}

\author{G.\ Escobedo}
\affiliation{Instituto Nacional de Astrof\'isica, \'Optica y Electr\'onica, Luis Enrique Erro 1, Tonantzintla, Puebla, C.P. 72840, M\'exico}

\author{A.\ Porras} 
\affiliation{Instituto Nacional de Astrof\'isica, \'Optica y Electr\'onica, Luis Enrique Erro 1, Tonantzintla, Puebla, C.P. 72840, M\'exico}

\author{E.\ Recillas}  
\affiliation{Instituto Nacional de Astrof\'isica, \'Optica y Electr\'onica, Luis Enrique Erro 1, Tonantzintla, Puebla, C.P. 72840, M\'exico}

\author{Y.~D.\ Mayya}
\affiliation{Instituto Nacional de Astrof\'isica, \'Optica y Electr\'onica, Luis Enrique Erro 1, Tonantzintla, Puebla, C.P. 72840, M\'exico}

\author{V.\ Chavushyan}
\affiliation{Instituto Nacional de Astrof\'isica, \'Optica y Electr\'onica, Luis Enrique Erro 1, Tonantzintla, Puebla, C.P. 72840, M\'exico}

\author{Beatriz Villarroel}
\affiliation{Nordita, KTH Royal Institute of Technology and Stockholm University, Roslagstullsbacken 23, SE-106 91 Stockholm, Sweden}
\affiliation{Instituto de Astrofisica de Canarias, Avda Via Lactea S/N, La Laguna, E-38205, Tenerife, Spain}

\author[0000-0002-8366-3373]{Zhongli Zhang}
\affiliation{Shanghai Astronomical Observatory, Chinese Academy of Sciences, Shanghai 200030, People's Republic of China}
\affiliation{Key Laboratory of Radio Astronomy, Chinese Academy of Sciences, 210033 Nanjing, Jiangsu, People's Republic of China}

\begin{abstract}
\noindent
We present the most extensive and well-sampled long-term multi-band near-infrared (NIR) temporal 
and spectral variability study of OJ 287, considered to be the best candidate binary supermassive black hole blazar. 
These observations were made between December 2007 and November 2021. The source underwent 
$\sim$ 2 -- 2.5 magnitude variations in the J, H, and Ks NIR bands. Over these long-term timescales 
there were no systematic trends in either flux or spectral evolution with time or with the 
source's flux states. However, on shorter timescales, there are significant variations in flux 
and spectra indicative of strong changes during different activity states. The NIR spectral energy 
distributions show diverse facets at each flux state, from the lowest to the highest. The spectra 
are, in general, consistent with a power-law spectral profile (within 10\%) and many of them 
indicate minor changes (observationally insignificant) in the shift of the peak. The NIR spectra generally steepens during bright phases.
We briefly discuss 
these behaviors in the context of blazar emission scenarios/mechanisms, OJ 287's well-known traditional 
behavior, and implications for models of the source central engine invoked for its long-term optical 
semi-periodic variations.
\end{abstract}


\section{Introduction}
\noindent
Blazars, referring to the union of BL Lacertae objects (BL Lacs) and
flat spectrum radio
quasars (FSRQs), is a subclass of radio-loud active galactic nuclei (AGNs)  
that host a large-scale relativistic jet of plasma pointing almost in our 
direction \citep{1995PASP..107..803U}. The jet is launched very near to the core 
formed by a central supermassive black hole (SMBH) of mass in the range of 
10$^{6}$ -- 10$^{10}$ M$_{\odot}$ and the plasma around it \citep{2002ApJ...579..530W}. 
Blazars are known for perennial 
dynamic variability, characterized by rapid and strong flux variations
in their emission that spans the entire electromagnetic spectrum from radio up to
$\gamma-$rays;  that emission
exhibits a broad bi-modal spectral energy distribution (SED) \citep{1998MNRAS.299..433F}.
The lower energy hump is attributed to synchrotron emission from relativistic leptons 
and the higher energy
hump to inverse Compton or hadronic processes
\citep[e.g.,][and references therein]{1983ApJ...264..296M,2003APh....18..593M,2017SSRv..207....5R}. \\
\\
Variability across the complete electromagnetic (EM) spectrum has been a key component in
the definition of blazars and is not only limited to flux but encompasses all the directly 
accessible observables. Blazar EM emission is predominantly 
non-thermal. In the absence of adequate spatial resolution, temporal flux variability is used 
to infer spatial scales of the emission region. Studies of the fluxes of blazars have
found them to be variable on almost all accessible timescales from the order of a few minutes
 to decades and more. In general, variability has been categorized into three subclasses:
intraday variability (IDV) focusing on variability over a day or 
less \citep{1989Natur.337..627M,1995ARA&A..33..163W}, short-term variability (STV) 
focusing on variability over days to several weeks, and long-term variability (LTV) 
focusing on timescales of months to years \citep{2004A&A...422..505G}. \\
\\
The BL Lac blazar OJ 287 ($\alpha_{2000.0} =$ 08h 54m 48.87s, $\delta_{2000.0} =$+20$^{\circ}$ 06$^{'}$ 30.$^{''}$64) 
is at  redshift $z =$ 0.306 \citep{1985PASP...97.1158S}. Optical observational data on this source actually date back to
1888 and using this century long light curve (LC), \citet{1988ApJ...325..628S}
noticed for the first time that the source appeared to show double-peaked 
outburst features which repeated with a 
period of $\sim$12 yrs. To explain this nominal quasi-periodic oscillation (QPO) feature 
in the long term optical LC, \citet{1988ApJ...325..628S} proposed a binary SMBH 
system for the blazar and predicted that the next double-peaked outburst would 
occur in 1994 $-$ 1995. An extensive global observing campaign called OJ-94 
was organized and the predicted double-peaked outbursts were really observed, 
with the second peak being detected $\sim$1.2 years after the first one
\citep{1996A&A...305L..17S,1996A&A...315L..13S}. The OJ-94 project supported the
basic model prediction but also revealed rather sharp rises of the predicted
flares, which led to a major modification of the  model, with the outbursts
now attributed to the impact of the secondary SMBH on the accretion disk of
the primary \citep{1996ApJ...460..207L}. Apart from this apparently well-established QPO, OJ 287 is the 
blazar with the highest number of claims of QPOs on a wide range of timescales, from a few
tens of minutes to decades and more across many EM bands \citep[e.g.][and references
therein]{1973ApJ...179..721V,1985Natur.314..146C,1985Natur.314..148V,1988ApJ...325..628S,
2016ApJ...832...47B,2018MNRAS.478.3199B,2020MNRAS.499..653K}.\\
\\
In the observing campaign of OJ 287 during 2005 -- 2007, the double-peaked outbursts
were detected respectively at the end of 2005 and end of 2007 i.e., separated
by $\sim$ 2 yr \citep{2009ApJ...698..781V}. For
the most recent predicted double-peaked outbursts, the first and second outbursts
were observed in December 2015 and July 2019, respectively, i.e., separated by $\sim$
3.5 yr \citep{2016ApJ...819L..37V,2017MNRAS.465.4423G,2020ApJ...894L...1L}. The
continued theoretical and observational efforts following this have led to better 
constraints on the timings of these outbursts and thus, the model as well. The latest iteration 
of the model incorporating improved treatment of dynamics with more physical aspects 
related to strong gravity and its consequences on the timing of the QPOs is presented 
in \citet{2018ApJ...866...11D}. Alternative interpretations of these recurrent outbursts 
invoke simple jet precession scenarios \citep[e.g.][and references therein]{2018MNRAS.478.3199B,2020Univ....6..191B}. 
The jet precession models, however, are not favored by the spectral changes reported in 
NIR to $\gamma-$ray during and after the most recent outbursts
of 2015 \citep{2017arXiv170802160O,2017IAUS..324..168K,Komo20,2018MNRAS.473.1145K,2018MNRAS.479.1672K,2020ApJ...890...47P} and 2019 \citep{Komo20,2021ApJ...921...18K,2022MNRAS.509.2696S}.
The timing of most recent outbursts (2015 and 2019) considered within the binary disk-impact model indicate a significant effect of gravitational wave (GW)  energy loss.
Detailed modeling suggests the rate of orbital shrinkage induced by GW emission
is $\sim 10^{-3}$ and has a non-negligible effect on the timing of this QPO \citep{2018ApJ...866...11D}. OJ 287 or other AGNs possessing close binary SMBHs are eventual candidates for direct detections of GW emission by the Pulsar Timing Array (PTA) or an interferometer in space {\citep[e.g.][]{2018MNRAS.481.2249C,2019A&ARv..27....5B,2019arXiv190706482B,2019BAAS...51c.123B}.} \\
\\
Early studies found OJ 287 to be the most dynamically variable BL Lac object,
exhibiting correlated multi-wavelength variability \citep[e.g.][and references therein]{1985PASP...97.1158S,1998A&AS..133..163F}.
In NIR bands, OJ 287 has been studied occasionally \citep{1992AJ....104...40T,1986Natur.324..546G,1984MNRAS.210..961H,1984MNRAS.211..497H}, 
but these studies normally have been limited to the duration of an ongoing enhanced activity period \citep{1986Natur.324..546G,2000A&AS..146..141P,2018MNRAS.473.1145K} and the  very few done over longer ($\gtrsim$ years)
duration mainly have very sparse data sampling \citep{2012ApJ...756...13B,2014A&A...562A..79S}. 
In the very first coordinated radio, NIR, and optical
monitoring, a $\rm \sim 25\%$ IDV variation at NIR was reported, slighly less than in
the optical \citep{1972ApJ...178L..51E}. In another study at NIR with UKIRT (United
Kingdom Infra Red Telescope), strong brightness variations in J, H, and K bands, along with some
unusual J-H and H-K color variations, were found \citep{1982MNRAS.201..479W}.
Motivated by this result, further monitoring in J-band with a temporal resolution of 5s revealed a
1 magnitude brightness change in 50s -- the fastest and strongest variation in any
BL Lac at that time. In a photo-polarimetric study during an outburst state in 1983,
strong variation in flux as well as polarization and an energy-dependent variation
in polarization was seen \citep{1984MNRAS.210..961H,1984MNRAS.211..497H}. Also, an 
excellent correlation between IR flux and spectral index, in the sense that as the source gets fainter
the spectrum gets steeper and vice versa, was found \citep{1986Natur.324..546G}. 
In 1993--1994, a continuous increase in NIR brightness was seen, with the maximum
brightness a factor of 3 higher since the start of monitoring. Smaller flares with an amplitude of up to one magnitude were seen on timescales of a few days \citep{1995A&AS..113..431K,2000A&AS..146..141P}. Though early studies are quite sparsely sampled
and limited at most to a few days, the compiled data show strong flux as well as
spectral variations with a brightness change of $\gtrsim 3.5$ mag between the
extremes at NIR bands i.e., by a factor of $\gtrsim 25$ in 
flux \citep[and references therein]{1994MNRAS.270..341L,1998A&AS..133..163F}.\\
\\
Later studies employing simultaneous NIR--optical data, much better sampled
than previous ones, and spanning over a few years timescales report magnitude variability
of around 2, or flux variations of $\gtrsim 6$ times, between the extremes, with the NIR changes slightly less than those in the optical
\citep{2012ApJ...756...13B,2014A&A...562A..79S}. 
Significant spectral changes at NIR energies
as well as a hysteresis between NIR and optical color variation have also been reported
\citep{2012ApJ...756...13B}. In terms of strong all around changes in observational
behavior, the period around the latest double-peaked outburst has been remarkable
\citep{2017MNRAS.465.4423G,2019AJ....157...95G,2018MNRAS.473.1145K,2021ApJ...921...18K,2022MNRAS.509.2696S}. 
However, the data used in these studies are mostly biased towards high activity states. \\
\\
The study of blazar variability on diverse timescales across the complete EM spectrum 
is one of the prominent areas of research in modern astronomy and astrophysics.
The NIR variability of blazars is comparatively less explored than many other
bands due to the paucity of NIR ground based telescopes. For building a NIR telescope,
one requires an observing site with low humidity, which most ground based 
observatories do not have. We have access to a 2.12 meter NIR telescope at an
excellent observing site in M{\'e}xico. We started a pilot project to study blazars'
temporal and spectral variabilities on diverse timescales in NIR bands in isolation
and/or with associated multi-wavelength observations. Under the project, here
we present the first densely sampled multi-band long-term  NIR temporal and spectral 
variability study of the blazar OJ 287 from 2007 December 18 -- 2021 November13. 
A multi-wavelength study reporting the spectral and temporal behavior will be presented 
in follow-up  work \citep[in preparation]{2022Kush} that will deal with the vastly different 
sampling and data integration times in different portions of the EM spectrum.\\
\\
In Section \S\ref{sec:obsdata}, we provide brief information about the observing facility, data acquisition, and 
reduction. In Section \S\ref{sec:result}, we present our results, and in Section \S\ref{sec:discuss} we give a discussion 
followed by a summary.

\section{Observations and Data Reduction}\label{sec:obsdata}
\noindent
The data of OJ287 used in this paper are part of the INAOE{\footnote{Instituto Nacional de Astrof\'{\i}sica, \'Optica y Electr\'onica, Mexico}}
NIR monitoring program of Blazars that started in 2005 (Carrasco, et al.\ in preparation) and have
been graciously provided by the members of the program. \\
\\
These new J, H, and Ks band NIR  photometric observations were obtained with the 2.12 meter telescope of the Guillermo 
Haro Astrophysical Observatory (OAGH) located in Cananea, Sonora, M{\'e}xico. The telescope is equipped with a NIR 
Camera named CANICA (the Cananea Near-Infrared Camera) which operates at multiple bands, including J (1.24 $\mu$m), 
H (1.63 $\mu$m) and Ks (2.12 $\mu$m) broad-bands.\\
\\
The camera is a 1024 pixel $\times$ 1024 pixel format HgCdTe Hawaii II array of 18.5 $\mu$m $\times$ 18.5 $\mu$m pixel 
size, covering a field of view of 5 arcmin $\times$ 5 arcmin for a plate scale 0.32 arcsec/pixel in the sky \citep{2017RMxAA..53..497C}. 
The frames were dark subtracted, flat fielded, and obtained at 7 dithered positions in the sky in a sequential manner
for the filters H, J, and Ks bands. Those frames were then median sky 
subtractedand finally, after shifting and registering, were co-added. Relative photometry is obtained for every 
co-added frame to the photometric values for point sources listed in the 2MASS (Two Micron All Sky Survey) in 
the field of view  of the camera.\\
\\ 
For OJ 287 the dithered images had typical exposure times of 30 sec, 
yielding total integration times of 210 sec for each filter. The number of comparison sources was typically 10.
In general, probable errors are 0.04, 0.03 and 0.04 magnitude in J, H, and Ks bands, respectively. The present 
data sample comprises $\sim$ 520 individual observations. These data, after correcting for redenning following 
\citet{1989ApJ...345..245C}, are reported in Table 1.

\startlongtable
\begin{deluxetable*}{cccccc}




\tablecaption{Reddening corrected NIR data of OJ 287 between 2007 -- 2021 (ref. \S\ref{sec:obsdata})}

\tablenum{1}

\tablehead{\colhead{JD} & \colhead{J} & \colhead{JD} & \colhead{H} & \colhead{JD} & \colhead{Ks} \\ 
\colhead{(2450000+)} & \colhead{(mag$\pm$error)} & \colhead{(2450000+)} & \colhead{(mag$\pm$error)} & \colhead{(2450000+)} & \colhead{(mag$\pm$error)} } 

\startdata
  4452.797360  & 12.105$\pm$0.03  &     4452.801526  & 11.276$\pm$0.03 & 4452.805692  & 10.942$\pm$0.03  \\
  4475.951106  & 11.798$\pm$0.04  &     4475.962679  & 11.045$\pm$0.03 & 4475.970317  & 10.301$\pm$0.03  \\
  4507.847707  & 12.179$\pm$0.03  &     4507.853436  & 11.307$\pm$0.02 & 4507.858968  & 10.636$\pm$0.04  \\
  4551.719372  & 12.704$\pm$0.03  &     4551.727011  & 11.844$\pm$0.05 & 4551.733261  & 11.005$\pm$0.02  \\
  4564.797741  & 13.021$\pm$0.04  &     4564.789940  & 12.150$\pm$0.03 & 4564.805738  & 11.270$\pm$0.02  \\
  4589.668243  & 12.850$\pm$0.05  &     4589.663381  & 12.040$\pm$0.05 & 4589.673104  & 11.166$\pm$0.01  \\
  4804.998881  & 12.627$\pm$0.05  &     4805.005826  & 11.658$\pm$0.06 & 4805.013466  & 10.842$\pm$0.03  \\
  4856.995965  & 12.542$\pm$0.03  &     4856.989390  & 11.638$\pm$0.06 & 4856.999784  & 10.689$\pm$0.03  \\
  4860.880016  & 12.718$\pm$0.02  &     4860.893905  & 11.750$\pm$0.03 & 4860.902932  & 10.866$\pm$0.03  \\
  4893.821950  & 12.240$\pm$0.03  &     4893.817714  & 11.272$\pm$0.05 & 4893.826394  & 10.421$\pm$0.03  \\
  4909.779554  & 13.002$\pm$0.03  &     4909.773652  & 12.179$\pm$0.04 & 4909.785109  & 11.184$\pm$0.06  \\
  4912.821687  & 12.917$\pm$0.05  &     4912.814049  & 11.997$\pm$0.05 & 4912.827937  & 11.025$\pm$0.05  \\
  4954.708825  & 13.388$\pm$0.05  &     4954.691466  & 12.497$\pm$0.05 &   --         &  --              \\
  4976.654647  & 13.474$\pm$0.01  &     4976.647703  & 12.494$\pm$0.06 & 4976.658119  & 11.616$\pm$0.03  \\
  5177.965478  & 12.137$\pm$0.03  &     5177.956449  & 11.227$\pm$0.05 & 5177.970339  & 10.396$\pm$0.04  \\
  5183.984137  & 12.232$\pm$0.03  &     5183.978685  & 11.325$\pm$0.06 & 5183.987737  & 10.607$\pm$0.04  \\
  5185.004181  & 12.142$\pm$0.03  &     5184.999551  & 11.370$\pm$0.06 & 5185.006264  & 10.619$\pm$0.05  \\
  5185.894403  & 12.189$\pm$0.03  &     5185.890352  & 11.291$\pm$0.02 & 5185.897991  & 10.522$\pm$0.05  \\
  5207.878302  & 12.091$\pm$0.09  &     5207.874830  & 11.233$\pm$0.07 & 5207.883858  & 10.471$\pm$0.08  \\
  5241.899977  & 12.293$\pm$0.06  &     5241.895533  & 11.421$\pm$0.04 & 5241.903450  & 10.707$\pm$0.02  \\
  5244.892790  & 12.226$\pm$0.02  &     5244.888160  & 11.476$\pm$0.06 & 5244.896158  & 10.897$\pm$0.07  \\
   --          &   --             &     5259.893016  & 11.702$\pm$0.06 &  --          &  --              \\
  5269.712567  & 12.709$\pm$0.03  &     5269.690346  & 11.905$\pm$0.06 & 5269.745203  & 11.298$\pm$0.09  \\
  5273.830616  & 12.741$\pm$0.05  &     5273.826866  & 11.940$\pm$0.03 & 5273.834366  & 11.303$\pm$0.09  \\
  5305.703974  & 13.234$\pm$0.02  &     5305.694947  & 12.247$\pm$0.04 & 5305.716472  & 11.616$\pm$0.11  \\
  5312.764392  & 13.432$\pm$0.04  &     5312.757448  & 12.604$\pm$0.03 & 5312.769947  & 11.937$\pm$0.12  \\
  5320.667784  & 13.175$\pm$0.01  &     5320.662924  & 12.367$\pm$0.05 & 5320.673339  & 11.603$\pm$0.02  \\
  5331.641018  & 13.486$\pm$0.06  &     5331.635463  & 12.638$\pm$0.03 & 5331.637546  & 11.829$\pm$0.07  \\
  5333.665130  & 13.788$\pm$0.04  &     5333.659876  & 12.907$\pm$0.06 & 5333.668047  & 12.193$\pm$0.02  \\
  5363.643642  & 13.324$\pm$0.01  &     5363.639892  & 12.533$\pm$0.04 & 5363.648433  & 12.636$\pm$0.19  \\
   --          &  --              &     5480.012163  & 12.576$\pm$0.01 &  --          &  --              \\
  5515.981656  & 12.885$\pm$0.04  &     5515.977489  & 11.991$\pm$0.01 & 5515.985615  & 11.213$\pm$0.09  \\
  5559.937581  & 13.059$\pm$0.03  &     5559.935405  & 12.513$\pm$0.01 & 5559.939711  & 11.514$\pm$0.03  \\
  5573.913484  & 12.838$\pm$0.03  &     5573.910127  & 12.006$\pm$0.07 & 5573.915822  & 11.232$\pm$0.04  \\
  5574.954757  & 12.574$\pm$0.04  &     5574.952245  & 11.824$\pm$0.07 & 5574.959965  & 11.167$\pm$0.05  \\
  5576.854051  & 12.767$\pm$0.06  &     5576.850683  & 11.957$\pm$0.04 & 5576.857130  & 11.094$\pm$0.07  \\
  5599.914132  & 12.733$\pm$0.07  &     5599.911829  & 11.916$\pm$0.05 & 5599.916493  & 11.016$\pm$0.04  \\
  5601.898171  & 12.795$\pm$0.04  &     5601.895613  & 11.894$\pm$0.04 & 5601.900463  & 11.044$\pm$0.06  \\
  5634.809086  & 13.180$\pm$0.02  &     5634.805822  & 12.351$\pm$0.04 & 5634.811655  & 11.681$\pm$0.04  \\
  5635.789826  & 13.128$\pm$0.03  &     5635.786736  & 12.317$\pm$0.03 & 5635.794549  & 11.587$\pm$0.04  \\
  5666.731273  & 13.126$\pm$0.02  &     5666.728565  & 12.304$\pm$0.06 & 5666.733773  & 11.391$\pm$0.07  \\
  5674.673090  & 13.183$\pm$0.04  &     5674.670093  & 12.538$\pm$0.04 & 5674.676609  & 11.757$\pm$0.02  \\
  5689.726944  & 13.106$\pm$0.03  &     5689.723935  & 12.260$\pm$0.02 & 5689.729919  & 11.494$\pm$0.08  \\
  5692.733819  & 13.202$\pm$0.05  &     5692.731273  & 12.396$\pm$0.04 & 5692.736343  & 11.812$\pm$0.07  \\
  5693.654502  & 13.146$\pm$0.05  &     5693.651944  & 12.341$\pm$0.07 & 5693.657384  & 11.641$\pm$0.07  \\
  5695.658877  & 13.072$\pm$0.09  &     5695.656389  & 12.260$\pm$0.04 & 5695.661424  & 11.493$\pm$0.11  \\
  5696.667708  & 13.057$\pm$0.07  &     5696.664583  & 12.257$\pm$0.04 & 5696.670671  & 11.470$\pm$0.07  \\
  5703.663021  & 13.191$\pm$0.03  &     5703.660729  & 12.401$\pm$0.07 & 5703.665046  & 11.659$\pm$0.07  \\
  6066.700995  & 12.238$\pm$0.07  &     6066.696111  & 11.446$\pm$0.07 & 6066.710139  & 10.714$\pm$0.04  \\
  6225.043171  & 13.086$\pm$0.09  &     6225.044977  & 12.306$\pm$0.09 & 6225.046632  & 11.903$\pm$0.09  \\
  6238.945556  & 12.869$\pm$0.07  &     6238.933958  & 12.365$\pm$0.05 & 6238.957002  & 11.406$\pm$0.08  \\
  6254.980648  & 12.891$\pm$0.04  &     6254.978009  & 12.011$\pm$0.06 & 6254.984063  & 11.410$\pm$0.09  \\
  6256.036019  & 13.143$\pm$0.04  &     6256.033646  & 12.349$\pm$0.06 & 6256.038472  & 11.825$\pm$0.04  \\
  6256.999641  & 13.483$\pm$0.03  &     6256.997130  & 12.532$\pm$0.05 & 6257.002269  & 12.039$\pm$0.06  \\
  6272.899988  & 13.369$\pm$0.03  &     6272.897095  & 12.523$\pm$0.05 & 6272.902824  & 11.890$\pm$0.09  \\
  6273.975336  & 13.594$\pm$0.05  &     6273.972315  & 12.673$\pm$0.04 & 6273.977882  & 11.898$\pm$0.07  \\
  6279.979630  & 13.629$\pm$0.05  &     6279.976354  & 12.707$\pm$0.05 & 6279.982813  & 12.020$\pm$0.12  \\
  6282.949213  & 13.587$\pm$0.03  &     6282.946111  & 12.766$\pm$0.07 & 6282.952465  & 12.294$\pm$0.08  \\
  6304.918600  & 13.824$\pm$0.04  &     6304.915382  & 12.944$\pm$0.03 & 6304.921910  & 12.256$\pm$0.04  \\
  6306.957569  & 13.714$\pm$0.04  &     6306.954745  & 12.832$\pm$0.05 & 6306.960278  & 12.204$\pm$0.05  \\
  6314.995185  & 13.825$\pm$0.02  &     6314.992731  & 12.912$\pm$0.06 & 6314.997662  & 12.514$\pm$0.05  \\
  6343.771204  & 12.733$\pm$0.04  &     6343.768623  & 11.925$\pm$0.04 & 6343.773796  & 11.346$\pm$0.11  \\
  6346.852350  & 12.596$\pm$0.05  &     6346.849653  & 11.745$\pm$0.06 & 6346.854815  & 11.020$\pm$0.08  \\
  6347.798472  & 12.890$\pm$0.04  &     6347.793623  & 11.906$\pm$0.05 & 6347.803472  & 11.422$\pm$0.04  \\
  6353.789502  & 12.847$\pm$0.05  &     6353.784005  & 11.932$\pm$0.04 & 6353.794259  & 11.493$\pm$0.03  \\
  6354.701076  & 12.856$\pm$0.05  &     6354.695799  & 11.945$\pm$0.04 & 6354.706400  & 11.459$\pm$0.06  \\
  6386.725046  & 13.151$\pm$0.03  &     6386.719734  & 12.335$\pm$0.03 & 6386.730370  & 11.845$\pm$0.06  \\
   --          &  --              &     6388.703796  & 12.358$\pm$0.04 &  --          &  --              \\
  6401.692975  & 13.557$\pm$0.05  &     6401.688738  & 12.637$\pm$0.04 & 6401.696609  & 11.950$\pm$0.07  \\
  6404.660370  & 13.191$\pm$0.04  &     6404.656933  & 12.471$\pm$0.02 & 6404.663924  & 11.783$\pm$0.03  \\
  6416.702060  & 12.977$\pm$0.06  &     6416.699630  & 12.043$\pm$0.05 & 6416.704421  & 11.449$\pm$0.09  \\
  6429.704456  & 12.935$\pm$0.06  &     6429.701238  & 12.054$\pm$0.06 & 6429.707951  & 11.541$\pm$0.07  \\
  6595.031354  & 13.145$\pm$0.04  &     6595.028958  & 12.663$\pm$0.03 & 6595.033322  & 11.798$\pm$0.04  \\
  6646.872130  & 13.428$\pm$0.06  &     6646.869850  & 12.663$\pm$0.05 & 6646.874329  & 11.866$\pm$0.12  \\
  6660.967743  & 13.143$\pm$0.06  &     6660.965532  & 12.245$\pm$0.05 & 6660.969722  & 11.786$\pm$0.06  \\
  6677.976331  & 13.450$\pm$0.08  &     6677.974039  & 12.650$\pm$0.03 & 6677.978530  & 11.810$\pm$0.03  \\
  6697.906204  & 13.059$\pm$0.02  &     6697.903854  & 12.272$\pm$0.03 & 6697.908773  & 11.567$\pm$0.05  \\
  6700.948924  & 12.630$\pm$0.06  &     6700.946435  & 11.760$\pm$0.05 & 6700.951042  & 10.962$\pm$0.01  \\
  6707.850625  & 12.381$\pm$0.04  &     6707.848333  & 11.535$\pm$0.08 & 6707.852581  & 10.626$\pm$0.05  \\
  6736.752569  & 12.946$\pm$0.03  &     6736.750590  & 11.986$\pm$0.08 & 6736.754664  & 11.107$\pm$0.04  \\
  6750.772593  & 12.603$\pm$0.07  &     6750.770799  & 11.733$\pm$0.04 & 6750.774259  & 10.891$\pm$0.06  \\
  6804.630799  & 13.670$\pm$0.08  &     6804.628183  & 12.671$\pm$0.07 & 6804.633009  & 11.807$\pm$0.09  \\
  6978.973391  & 13.668$\pm$0.07  &     6978.970498  & 12.749$\pm$0.07 & 6978.976354  & 11.989$\pm$0.06  \\
  6993.009502  & 13.587$\pm$0.02  &     6993.006250  & 12.741$\pm$0.04 & 6993.012986  & 11.798$\pm$0.03  \\
  7007.018704  & 12.792$\pm$0.04  &     7007.013889  & 12.029$\pm$0.03 & 7007.023738  & 11.268$\pm$0.06  \\
  7021.041123  & 12.849$\pm$0.03  &     7021.039155  & 12.011$\pm$0.05 & 7021.043090  & 11.289$\pm$0.03  \\
  7032.954213  & 12.792$\pm$0.02  &     7032.951887  & 12.004$\pm$0.07 & 7032.956667  & 11.240$\pm$0.02  \\
  7035.938623  & 12.839$\pm$0.02  &     7035.936250  & 12.018$\pm$0.02 & 7035.941146  & 11.323$\pm$0.03  \\
  7079.808021  & 13.038$\pm$0.03  &     7079.804618  & 12.230$\pm$0.03 & 7079.812130  & 11.561$\pm$0.05  \\
  7081.823090  & 13.098$\pm$0.05  &     7081.820509  & 12.558$\pm$0.05 & 7081.825799  & 11.637$\pm$0.05  \\
  7095.869248  & 12.947$\pm$0.03  &     7095.866748  & 12.181$\pm$0.05 & 7095.871910  & 11.251$\pm$0.04  \\
  7112.763403  & 12.991$\pm$0.07  &     7112.761563  & 12.191$\pm$0.04 & 7112.765347  & 11.502$\pm$0.04  \\
  7121.772153  & 13.108$\pm$0.09  &     7121.769063  & 12.159$\pm$0.03 & 7121.774502  & 11.493$\pm$0.08  \\
  7140.703819  & 12.891$\pm$0.04  &     7140.700602  & 12.113$\pm$0.03 & 7140.707350  & 11.405$\pm$0.04  \\
  7154.708553  & 12.759$\pm$0.03  &     7154.705637  & 11.945$\pm$0.04 & 7154.711435  & 11.275$\pm$0.04  \\
  7169.656632  & 13.288$\pm$0.08  &     7169.653553  & 12.276$\pm$0.06 & 7169.659340  & 11.794$\pm$0.04  \\
  7332.964664  & 13.185$\pm$0.03  &     7332.960799  & 12.316$\pm$0.03 & 7332.968646  & 11.419$\pm$0.03  \\
  7348.034618  & 13.017$\pm$0.06  &     7348.032303  & 12.173$\pm$0.03 & 7348.036887  & 11.444$\pm$0.03  \\
  7362.930382  & 11.978$\pm$0.03  &     7362.927535  & 11.115$\pm$0.04 & 7362.933252  & 10.373$\pm$0.03  \\
  7365.024826  & 11.768$\pm$0.03  &     7365.021296  & 10.929$\pm$0.04 & 7365.027350  & 10.044$\pm$0.04  \\
  7365.971898  & 12.054$\pm$0.03  &     7365.968808  & 11.188$\pm$0.03 & 7365.974838  & 10.495$\pm$0.03  \\
  7373.960440  & 12.347$\pm$0.05  &     7373.955752  & 11.547$\pm$0.03 & 7373.964803  & 10.696$\pm$0.04  \\
  7375.011076  & 12.606$\pm$0.03  &     7375.008565  & 11.659$\pm$0.05 & 7375.013438  & 10.997$\pm$0.05  \\
  7376.013507  & 12.450$\pm$0.05  &     7376.006389  & 11.515$\pm$0.04 & 7376.020926  & 10.822$\pm$0.06  \\
  7378.038900  & 12.318$\pm$0.04  &     7378.035880  & 11.424$\pm$0.04 & 7378.041516  & 10.704$\pm$0.06  \\
  7398.004931  & 12.806$\pm$0.04  &     7398.001829  & 12.082$\pm$0.06 & 7398.007650  & 11.247$\pm$0.06  \\
  7414.942199  & 13.232$\pm$0.05  &     7414.939271  & 12.322$\pm$0.05 & 7414.947569  & 11.626$\pm$0.04  \\
  7415.920486  & 13.381$\pm$0.05  &     7415.917326  & 12.580$\pm$0.02 & 7415.923889  & 11.817$\pm$0.04  \\
  7417.925231  & 13.424$\pm$0.02  &     7417.922303  & 12.654$\pm$0.03 & 7417.928264  & 11.817$\pm$0.03  \\
  7418.876319  & 13.578$\pm$0.03  &     7418.873403  & 12.664$\pm$0.04 & 7418.879294  & 11.913$\pm$0.03  \\
  7433.918194  & 12.006$\pm$0.03  &     7433.915289  & 11.245$\pm$0.05 & 7433.921042  & 10.495$\pm$0.05  \\
  7435.834387  & 12.088$\pm$0.04  &     7435.830475  & 11.228$\pm$0.03 & 7435.837326  & 10.521$\pm$0.03  \\
  7436.826690  & 12.143$\pm$0.04  &     7436.823947  & 11.302$\pm$0.04 & 7436.829780  & 10.569$\pm$0.03  \\
  7441.708877  & 12.253$\pm$0.04  &     7441.704965  & 11.447$\pm$0.05 & 7441.712477  & 10.572$\pm$0.04  \\
  7444.836910  & 12.197$\pm$0.04  &     7444.833611  & 11.405$\pm$0.04 & 7444.854340  & 10.583$\pm$0.04  \\
  7447.826991  & 11.948$\pm$0.06  &     7447.824155  & 10.949$\pm$0.04 & 7447.829838  & 10.283$\pm$0.06  \\
  7448.936030  & 11.835$\pm$0.08  &     7448.933403  & 11.168$\pm$0.03 & 7448.938426  & 10.923$\pm$0.06  \\
  7466.799398  & 12.407$\pm$0.05  &     7466.797257  & 11.529$\pm$0.04 & 7466.801898  & 10.900$\pm$0.07  \\
  7481.836609  & 12.555$\pm$0.05  &     7481.832928  & 11.748$\pm$0.03 & 7481.840324  & 10.839$\pm$0.04  \\
  7493.745822  & 12.729$\pm$0.04  &      --          &  --             & 7493.748507  & 11.177$\pm$0.04  \\
  7495.712326  & 12.816$\pm$0.06  &     7495.709433  & 11.860$\pm$0.04 & 7495.715336  & 11.017$\pm$0.05  \\
  7496.726366  & 12.628$\pm$0.05  &     7496.723634  & 11.828$\pm$0.04 & 7496.729167  & 11.151$\pm$0.03  \\
  7497.674780  & 12.762$\pm$0.05  &     7497.671574  & 11.979$\pm$0.06 & 7497.678252  & 11.191$\pm$0.05  \\
  7688.011887  & 11.771$\pm$0.02  &     7688.009560  & 10.996$\pm$0.02 & 7688.014572  & 10.175$\pm$0.04  \\
  7689.016863  & 11.685$\pm$0.03  &     7689.014410  & 11.186$\pm$0.03 & 7689.019595  & 10.263$\pm$0.04  \\
  7689.987280  & 11.989$\pm$0.04  &     7689.984850  & 11.066$\pm$0.03 & 7689.990081  & 10.413$\pm$0.07  \\
  7706.036991  & 11.966$\pm$0.05  &     7706.034711  & 11.218$\pm$0.03 & 7706.039618  & 10.622$\pm$0.05  \\
  7761.922882  & 12.244$\pm$0.05  &     7761.919039  & 11.343$\pm$0.05 & 7761.925336  & 10.651$\pm$0.06  \\
  7764.949549  & 12.146$\pm$0.04  &     7764.946343  & 11.407$\pm$0.05 & 7764.952697  & 10.689$\pm$0.04  \\
  7771.945058  & 12.453$\pm$0.05  &     7771.942164  & 11.682$\pm$0.05 & 7771.948032  & 10.891$\pm$0.04  \\
  7787.826586  & 12.221$\pm$0.03  &     7787.823808  & 11.349$\pm$0.03 & 7787.830648  & 10.695$\pm$0.09  \\
  7789.911285  & 11.965$\pm$0.03  &     7789.908889  & 11.197$\pm$0.03 & 7789.913426  & 10.505$\pm$0.07  \\
  7805.881574  & 12.336$\pm$0.04  &     7805.878796  & 11.465$\pm$0.04 & 7805.884873  & 10.884$\pm$0.04  \\
  7816.843542  & 12.338$\pm$0.02  &     7816.835741  & 11.626$\pm$0.03 & 7816.840972  & 10.796$\pm$0.04  \\
  7819.820799  & 12.464$\pm$0.04  &     7819.817894  & 11.674$\pm$0.02 & 7819.823704  & 10.937$\pm$0.04  \\
  7827.787164  & 12.296$\pm$0.02  &     7827.785694  & 11.551$\pm$0.05 & 7827.793519  & 10.779$\pm$0.02  \\
  7867.683866  & 13.316$\pm$0.05  &     7867.682940  & 12.380$\pm$0.03 & 7867.684958  & 11.730$\pm$0.04  \\
  7878.738183  & 12.709$\pm$0.04  &     7878.735405  & 12.022$\pm$0.04 & 7878.741481  & 11.277$\pm$0.03  \\
  8118.848299  & 12.734$\pm$0.06  &     8118.841991  & 11.671$\pm$0.03 & 8118.855289  & 11.068$\pm$0.05  \\
  8140.908125  & 12.959$\pm$0.07  &     8140.904005  & 12.053$\pm$0.03 & 8140.914340  & 11.298$\pm$0.06  \\
  8146.955255  & 12.654$\pm$0.04  &     8146.950729  & 11.783$\pm$0.03 & 8146.959248  & 10.834$\pm$0.07  \\
  8198.785648  & 13.090$\pm$0.04  &     8198.780880  & 12.240$\pm$0.04 & 8198.787963  & 11.518$\pm$0.03  \\
  8204.733993  & 13.100$\pm$0.04  &     8204.732500  & 12.131$\pm$0.04 & 8204.735880  & 11.191$\pm$0.07  \\
  8244.759329  & 13.005$\pm$0.03  &     8244.751019  & 12.212$\pm$0.03 & 8244.765243  & 11.396$\pm$0.03  \\
  8257.641424  & 13.353$\pm$0.08  &     8257.636748  & 12.251$\pm$0.04 &  --          &  --              \\
  8448.052188  & 13.321$\pm$0.05  &     8448.048657  & 12.540$\pm$0.04 & 8448.056215  & 11.604$\pm$0.05  \\
  8540.875486  & 13.557$\pm$0.05  &     8540.868669  & 12.692$\pm$0.02 & 8540.882789  & 12.064$\pm$0.03  \\
  8571.836053  & 13.390$\pm$0.06  &     8571.829630  & 12.660$\pm$0.04 & 8571.839583  & 11.838$\pm$0.04  \\
  8575.706343  & 13.236$\pm$0.05  &     8575.703264  & 12.495$\pm$0.02 & 8575.707627  & 11.771$\pm$0.06  \\
  8582.743125  & 13.254$\pm$0.03  &     8582.737755  & 12.477$\pm$0.03 & 8582.754028  & 11.796$\pm$0.02  \\
  8602.656944  & 13.592$\pm$0.03  &     8602.655394  & 12.838$\pm$0.02 & 8602.665301  & 11.981$\pm$0.04  \\
  8603.718380  & 13.572$\pm$0.02  &     8603.711111  & 12.830$\pm$0.03 & 8603.718056  & 12.094$\pm$0.03  \\
  8612.695833  & 13.550$\pm$0.03  &     8612.696748  & 12.770$\pm$0.04 & 8612.702616  & 11.952$\pm$0.05  \\
  8793.024757  & 12.790$\pm$0.03  &     8793.027627  & 11.913$\pm$0.03 & 8793.030104  & 11.239$\pm$0.04  \\
  8835.048819  & 13.499$\pm$0.02  &     8835.042002  & 12.719$\pm$0.03 & 8835.056366  & 11.842$\pm$0.03  \\
  8836.952037  & 13.448$\pm$0.04  &     8836.945127  & 12.592$\pm$0.03 &  --          &  --              \\
  8856.916563  & 12.925$\pm$0.03  &     8856.905081  & 12.153$\pm$0.05 & 8856.921759  & 11.375$\pm$0.05  \\
   --          &  --              &     8863.001574  & 12.645$\pm$0.06 &  --          &  --              \\
  8866.942558  & 13.043$\pm$0.04  &     8866.938056  & 12.352$\pm$0.04 & 8866.947674  & 11.474$\pm$0.03  \\
  8881.957072  & 13.056$\pm$0.03  &     8881.953681  & 12.666$\pm$0.06 & 8881.960440  & 11.831$\pm$0.06  \\
  8882.899502  & 13.105$\pm$0.03  &     8882.893669  & 12.328$\pm$0.04 & 8882.905775  & 11.594$\pm$0.06  \\
  8884.899606  & 13.115$\pm$0.02  &     8884.895359  & 12.413$\pm$0.03 & 8884.898171  & 11.639$\pm$0.03  \\
  8886.837674  & 13.259$\pm$0.03  &     8886.832407  & 12.412$\pm$0.04 & 8886.840556  & 11.788$\pm$0.04  \\
  8893.913889  & 13.380$\pm$0.03  &     8893.908657  & 12.659$\pm$0.04 & 8893.914271  & 11.838$\pm$0.04  \\
  8913.878137  & 13.207$\pm$0.03  &     8913.867593  & 12.350$\pm$0.04 & 8913.881528  & 11.519$\pm$0.05  \\
  8928.769537  & 12.766$\pm$0.05  &     8928.766505  & 11.987$\pm$0.03 & 8928.774757  & 11.230$\pm$0.04  \\
  9221.020752  & 12.674$\pm$0.05  &     9221.014132  & 11.715$\pm$0.04 & 9221.028171  & 10.984$\pm$0.03  \\
  9267.925313  & 12.981$\pm$0.08  &     9267.917847  & 12.149$\pm$0.05 & 9267.933484  & 11.296$\pm$0.06  \\
  9307.813889  & 12.817$\pm$0.04  &     9307.809769  & 11.979$\pm$0.05 & 9307.826227  & 11.153$\pm$0.05  \\
  9308.819769  & 12.572$\pm$0.05  &     9308.813831  & 11.820$\pm$0.04 & 9308.825694  & 11.114$\pm$0.06  \\
  9326.738912  & 13.472$\pm$0.08  &     9326.736539  & 12.440$\pm$0.04 & 9326.742685  & 11.831$\pm$0.04  \\
  9342.718310  & 13.610$\pm$0.05  &     9342.714676  & 12.813$\pm$0.05 & 9342.721910  & 12.098$\pm$0.06  \\
  9356.686979  & 13.136$\pm$0.04  &     9356.675532  & 12.226$\pm$0.04 & 9356.694931  & 11.443$\pm$0.04  \\
  9359.721157  & 12.961$\pm$0.03  &     9359.715139  & 12.215$\pm$0.03 &  --          &  --              \\
  9367.672188  & 13.483$\pm$0.07  &     9367.669433  & 12.510$\pm$0.04 & 9367.676481  & 11.521$\pm$0.07  \\
  9369.638414  & 12.971$\pm$0.02  &     9369.632141  & 12.184$\pm$0.03 & 9369.639965  & 11.399$\pm$0.03  \\
  9382.641609  & 13.608$\pm$0.06  &     9382.637188  & 12.669$\pm$0.07 & 9382.647118  & 11.861$\pm$0.07  \\
  9532.018325  & 12.995$\pm$0.07  &     9532.014527  & 12.050$\pm$0.04 & 9532.022374  & 11.211$\pm$0.04  \\
\enddata




\end{deluxetable*}

\begin{figure*}[!ht]
\centering
\includegraphics[scale=1.3]{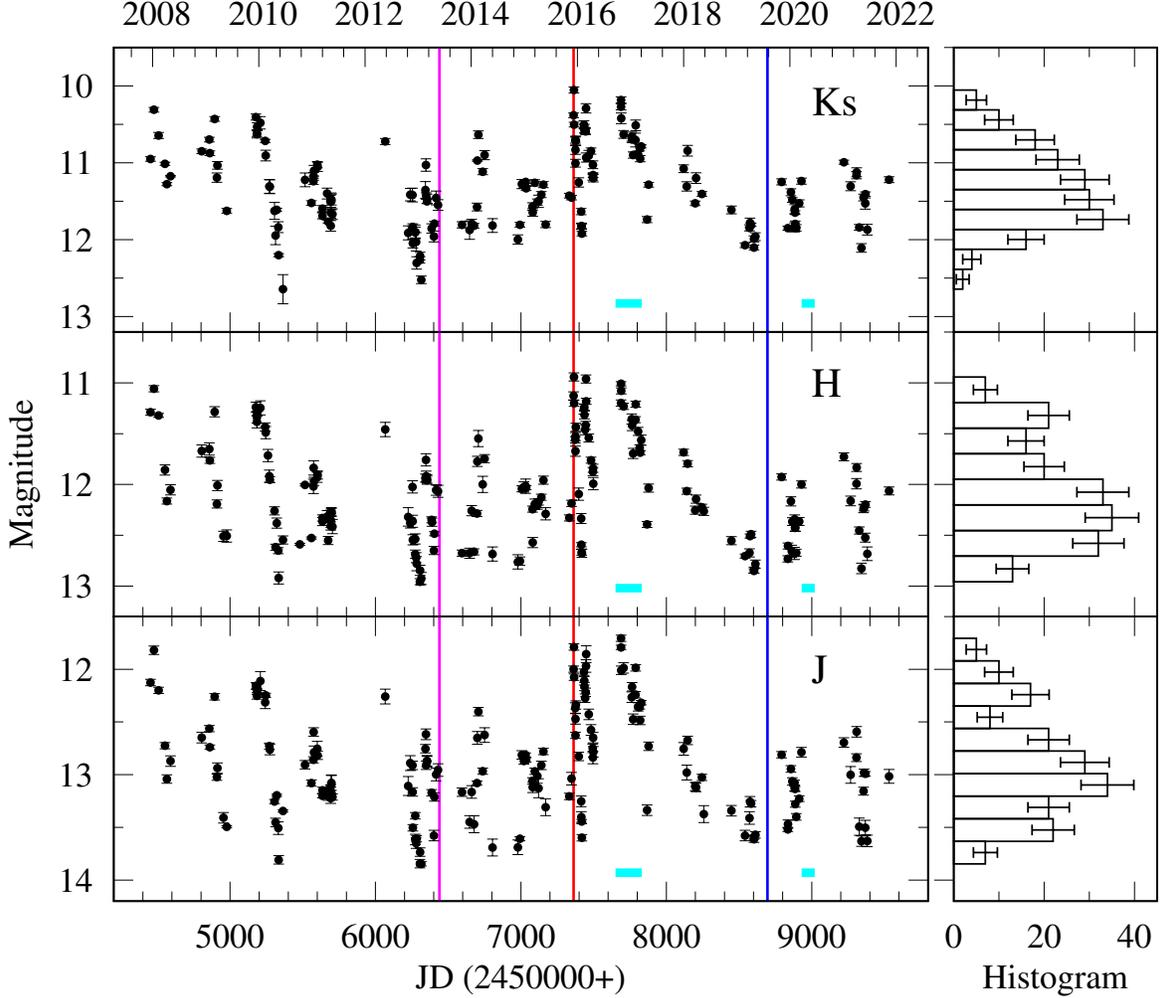}
\caption{Multi-band NIR variability light curves of the blazar OJ 287 during 2007 December -- 2021 November. From 
bottom to top, the left panels show J, H, and Ks calibrated magnitudes, respectively. The right panels show how 
many measurements fall into each equal bin, the widths of which are assigned through the Knuth method \citep{2006physics...5197K} 
and differ slightly for each band. The magenta, red, and blue vertical lines, respectively, mark the 
first sighting of a rather sharp NIR-optical spectral break in end May 2013 and the flux peaks of the double-peaked outbursts 
of the $\sim 12$-yr QPO seen in end 2015 and mid 2019. The horizontal cyan lines mark the durations of the brightest X-ray activity phases as reported in the literature.}
\label{fig:nirlc}
\end{figure*}

\section{Results}\label{sec:result}

\noindent
In Figure 1 we present the J, H, and Ks band NIR photometric light curves (LCs)
generated from our new observations taken during 2007 December -- 2021 November.
This is the most extensive and well-sampled long-term NIR photometric study of
the blazar OJ 287. On visual inspection the J, H, and Ks band LCs all clearly
show large amplitude flux variations. Several substantial flaring events in the
photometric observations in all three bands are seen. In the following subsections, 
we discuss the NIR temporal and spectral variability properties of the blazar OJ
287 on LTV timescales.

\subsection{LC Analysis Techniques}

\noindent
To calculate the amplitude of LTV variability and interband cross correlations in the NIR J, H, and Ks
bands, the methods we used are briefly described below.  

\subsubsection{Amplitude of Variability}

\noindent
The percentage of the amplitude of the variability in magnitude (and color) on 
LTV timescales is described by the  parameter, $A$, which  can be defined using the following equation introduced by 
\citet{1996A&A...305...42H} 

\begin{eqnarray}
A = 100\times \sqrt{{(A_{max}-A_{min}})^2 - 2\sigma^2}(\%) 
\end{eqnarray} 

\noindent
Here $A_{max}$ and $A_{min}$ are the maximum and minimum values, respectively, in the 
calibrated magnitude or color of the LC of the blazar, and $\sigma$ is the mean measurement error. 

\subsubsection{Discrete Cross-correlation Function}
\noindent
We carried out the cross-correlation analysis between the NIR bands using the
z-transformed Discrete Cross-correlation \citep[zDCF;][]{1997ASSL..218..163A,
2013arXiv1302.1508A} method. It is broadly similar to the traditional DCF except
that the correlation coefficient errors are estimated using the z-transform, given by
\begin{equation}
 z = \frac{1}{2} ln\left(\frac{1+r}{1-r}\right), 
 \zeta = \frac{1}{2} ln\left(\frac{1+\rho}{1-\rho}\right), 
 r = tanh(z), \nonumber
 \end{equation}
where $r$ and $\rm \rho$  represent the bin correlation coefficient and the unknown
population correlation coefficient, respectively. The correlation coefficients are
estimated by constructing all possible time lag data pairs ($x_i$,
$y_i$) between the two light curves as
\begin{equation}
 r = \frac{\sum_i^{n}(x_i - \overline{x})(y_i - \overline{y})}{\sigma_x \sigma_y}, 
 s_x^2 = \frac{1}{n-1} \sum_i^{n}(x_i - \overline{x})^2. \nonumber
 \end{equation}
In order to obtain the mean and variance of $z,  \rho=r$ is assumed \citep[][]{2013arXiv1302.1508A}. The reason for making 
the z-transformation is that the correlation coefficients
are not normally distributed in the real space. This method
is applicable to both uniformly and sparse, non-uniformly, sampled time series data.
It employs Fisher's z-transform and equal population binning to handle the bias arising
due to sampling and skewness and fares better compared to 
the traditional approaches \citep{1997ASSL..218..163A,2013arXiv1302.1508A}.
The errors were estimated using the Monte Carlo method
by simulating 1000 pairs of light curves from the observed light curves by adding a Gaussian
noise extracted from the measured error bars. The resulting cross-correlation
results are shown in Figure 2. The peaks at zero lag signify that the multi-band
NIR variations are simultaneous.

\subsection{Long Term Variability}
\noindent
Our typical observational cadence of once a month, with a daily follow-up around
the higher activity phases, allow us to explore long-term variations of OJ 287 in multi-band 
NIR flux, color, spectral index, and spectral energy distributions. We also discuss
the detection of a large number of flaring events during the whole observing duration.  

\subsubsection{Flux Variability}

\begin{deluxetable}{cccc}
\tablenum{2}
\tablecaption{Results of LTV Flux Variations \label{tab:messier}}
\tablewidth{0pt}
\tablehead{
\colhead{Band} & \colhead{Duration} & \colhead{Variable} & \colhead{A(\%)}
}
\startdata
J  & 2007-12-18  $-$ 2021-11-13  & Var & 213.9 \\
H  & 2007-12-18  $-$ 2021-11-13  & Var & 201.4 \\
Ks & 2007-12-18  $-$ 2021-11-13  & Var & 259.1 \\
\enddata
\end{deluxetable}

\noindent
Large amplitude significant flux variability from OJ 287 on LTV timescales is clearly visible 
from the three panels of Figure 1, where the J, H, and Ks band LCs are presented from bottom 
to top panels, respectively. We have calculated the variability amplitudes in the J, 
H, and Ks NIR photometric bands, and the results are reported in Table 2. We found the faintest
level of the blazar in J, H, and Ks bands were 13.846 mag at JD 2456314.995185, 12.957 mag at
JD 2456304.915382, and 12.645 mag at JD 2455363.648433, respectively. Similarly the observed 
brightest levels are 11.706 mag at JD 2457689.016863, 10.942 mag at JD 2457365.021296, and 
10.053 mag at JD 2457365.027350, in the J, H, and Ks bands, respectively. 
In terms of fluxes, the amplitudes of variation given in Table 2 correspond to changes by a
factor of roughly 7.2, 6.4, and 10.9 in the J, H, and Ks bands respectively. In the nearly 
14 years long NIR observational duration, the large amplitude variations in the blazar 
LCs indicate that we have observed in the source in low, intermediate, high, and
possibly even outburst, flux states. Historically, the brightest reported NIR magnitudes of 
OJ 287 were  J = 10.73 mag, H = 9.94 mag, K = 8.81 mag and the faintest, J = 14.60 mag, H = 13.73 mag, 
K = 12.75 mag \citep{1998A&AS..133..163F}. If we compare them with our data presented here, it 
is clearly seen that we have observed the blazar right in between its historically brightest 
and faintest states. \\
\\
Visually, it appears from Figure 1 that the J, H, Ks NIR bands follow the same variability 
pattern. To further examine the variability relations between these NIR bands, we performed DCF 
analyses using the zDCF method between these bands as shown in Figure 2. Strong correlations
with zero lag are found in the different combination of all three NIR bands. These correlations
strongly indicate that the emission in J, H, and Ks bands are cospatial and emitted from the same 
population of leptons.    

\begin{figure}[ht!]
\centering
\includegraphics[scale=0.7]{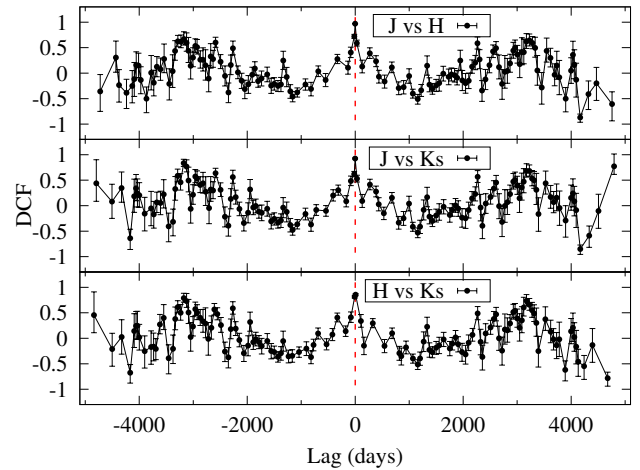}
\caption{DCF plots using the zDCF method between NIR J, H, and Ks bands for the total duration of observations. The time lag and DCF values are given on the X-axis and Y-axis, respectively. \label{fig:zdcf}}
\end{figure}

\subsubsection{Color Variability}
\begin{deluxetable}{ccccc}
\tablenum{3}
\tablecaption{Color Variation with Respect to Time on LTV\label{tab:colTime}}
\tablewidth{0pt}
\tablehead{
\colhead{Color}   & \colhead{$\rm m_2 (\times 10^{-6})$} & \colhead{$\rm c_2 $} & \colhead{$\rm r_2$} & \colhead{$\rm p_2$} \\
\colhead{Indices} & & & &
}
\startdata
J -- H  & $-$11.3$\rm\pm$5.9 & ~~~28$\rm\pm$14    & ~~0.017 & {0.06} \\
J -- Ks & $-$11.3$\rm\pm$8.0 & ~~~29$\rm\pm$19    & ~~0.006 & {0.16} \\
H -- Ks & ~~~~0.31$\rm\pm$7.61 & ~~~0$\rm\pm$18    & $-$0.006 & {0.97} \\
\enddata

\vspace*{0.2cm}
\noindent
{\bf Note:} m$_{2}$ = slope and c$_{2}$ = intercept of color against H mag; 
r$_{2}$ = coefficient of determination ($\rm R^2$); p$_{2}$ (0.05) = null hypothesis rejection probability.

\end{deluxetable}

\begin{deluxetable}{ccccc}
\tablenum{4}
\tablecaption{Color Variation with Respect to H-band magnitude on LTV\label{tab:colMag}}
\tablewidth{0pt}
\tablehead{
\colhead{Color} & \colhead{$\rm m_2$} & \colhead{$\rm c_2$} & \colhead{$\rm r_2$} & \colhead{$\rm p_2$} \\
\colhead{Indices} & & & &
}
\startdata
J -- H  & -0.026 $\rm\pm$0.015 & 1.14$\rm\pm$0.18 &  0.013 & 0.08\\
J -- Ks & -0.018 $\rm\pm$0.022 & 1.80$\rm\pm$0.26 & -0.002 & 0.41\\
H -- Ks & 0.028  $\rm\pm$0.020 & 0.41$\rm\pm$0.24 &  0.007 & 0.16\\
\enddata

\vspace*{0.2cm}
\noindent
{\bf Note:} m$_{2}$ = slope and c$_{2}$ = intercept of color against H mag; 
r$_{2}$ = coefficient of determination ($\rm R^2$); p$_{2}$ (0.05) = null hypothesis rejection probability.
\end{deluxetable}

\begin{figure*}[ht!]
\centering
\includegraphics[scale=1.3]{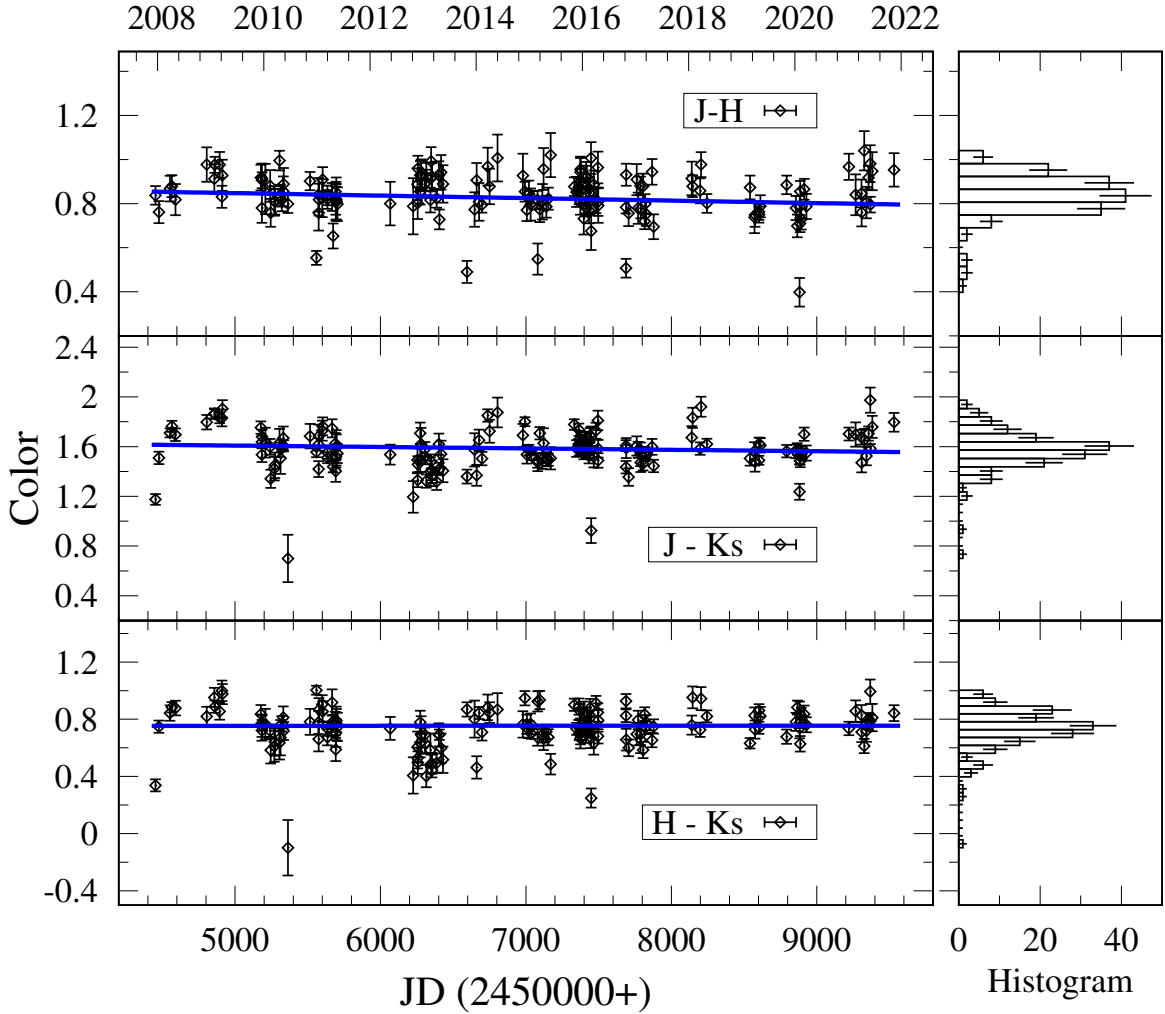}
\caption{As in Fig.\ 1 for the NIR color variability for the entire duration of these observations of OJ 287. The panels on the right show the spectral index histogram. \label{fig:tVsCor}}
\end{figure*}

\begin{figure*}[ht!]
\centering
\includegraphics[scale=1.3]{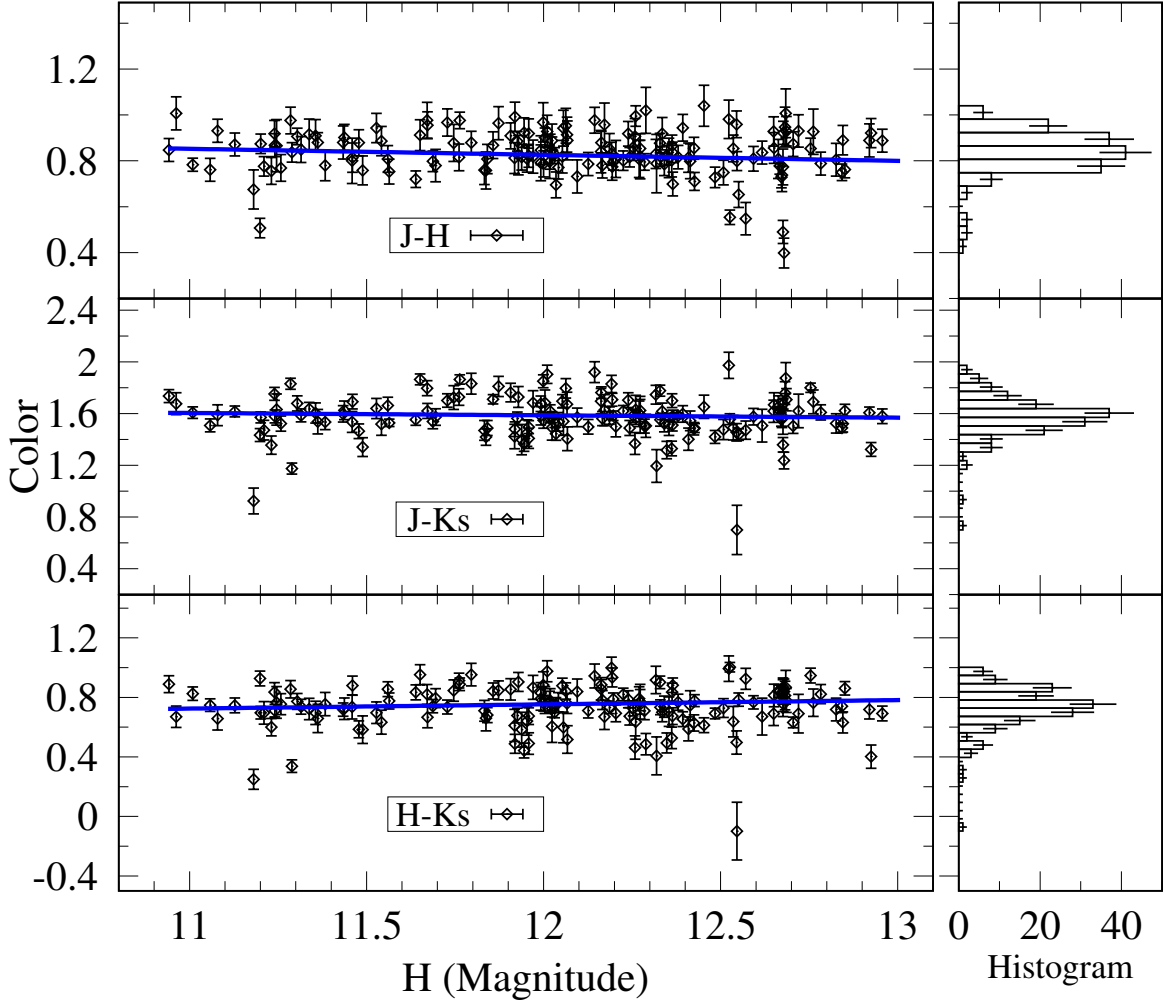}
\caption{As in Fig.\ 3 for NIR color--magnitude plots for OJ 287.  \label{fig:HmagVsCor}}
\end{figure*}

\noindent
For the total duration of our observations of OJ 287, NIR color variations with respect 
to time (color vs.\ time) and with respect to H-band magnitude (color vs.\ 
magnitude) are displayed in Figures 3 and 4, respectively. On visual inspection
both figures show weak evidence of color variations, but there are no 
consistent systematic trends in the color variations with respect to time or
H-band magnitude. To further examine the color variation, we did straight line 
fits to the color versus time, and color versus H-mag, plots in Figure 3 and 
Figure 4, respectively. The straight line fit parameters values e.g., the slopes, 
$m$, the intercepts, $c$, the linear Pearson correlation coefficients, $r$, and 
the corresponding null hypothesis rejection probability, $p$, for color versus time and 
color versus H-band magnitude are given in Tables 3 and 4, respectively.

\subsubsection{Spectral Index Variations and SEDs}

\begin{figure*}[ht!]
\centering
\includegraphics[scale=1.3]{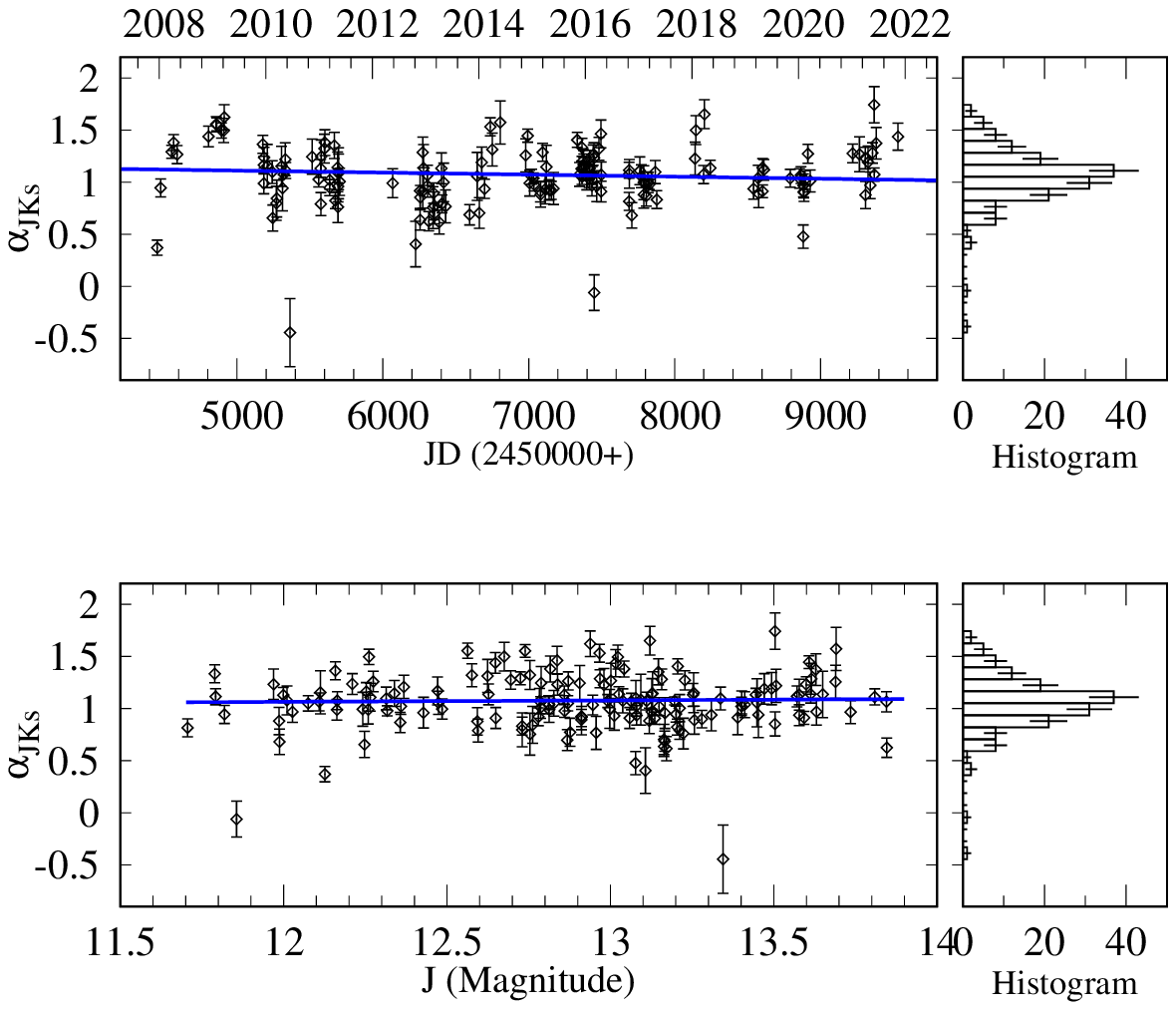}
\caption{NIR spectral index variation with time and J-band magnitude covering the entire observation period of OJ 287. The panels on the right show the spectral index histogram. \label{fig:indx}}
\end{figure*}

\begin{figure*}
\begin{interactive}{animation}{SEDmovie.mp4}
\plotone{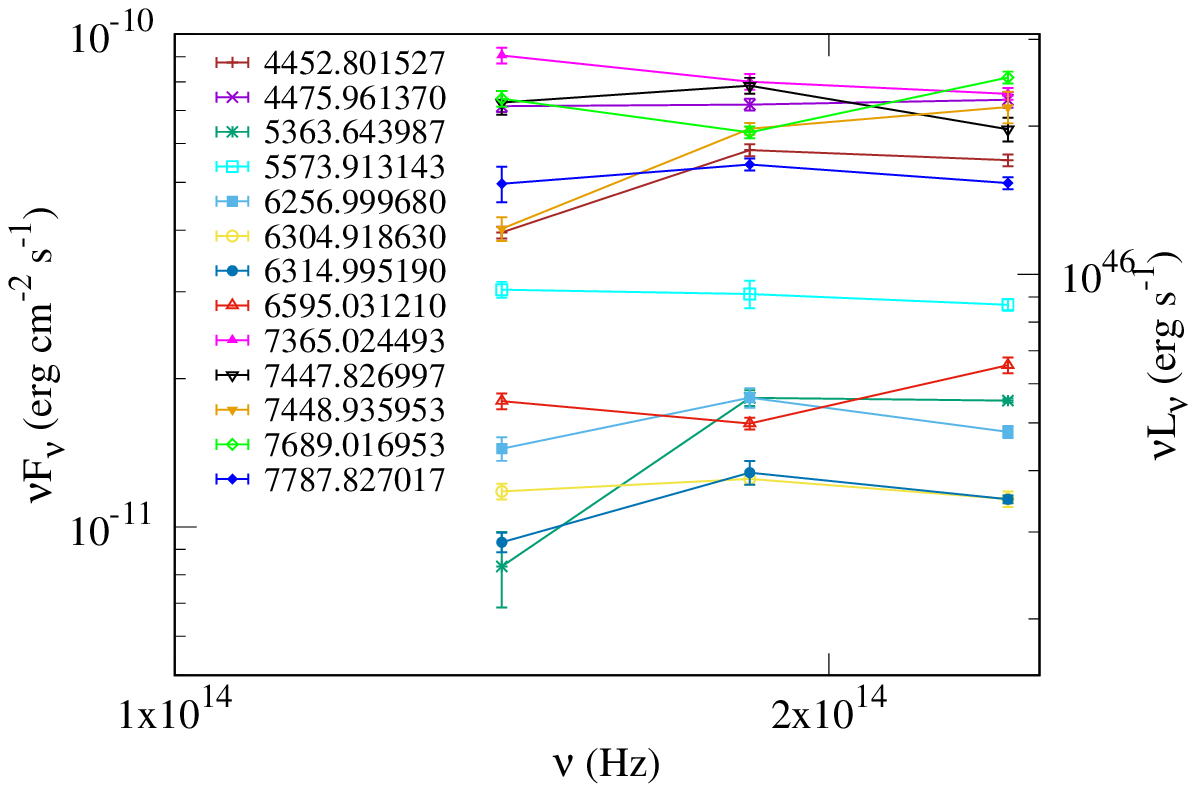}
\end{interactive}
\caption{Plot showing a glimpse of diverse NIR spectral phases of OJ 287. The accompanying video shows the NIR SED evolution with time. The video duration is 17 seconds. 
\label{fig:video}}
\end{figure*}

\noindent
In these magnitude measurements the color variations encode spectral information across the NIR. 
Making the assumption of a power-law spectrum across these bands we find
\begin{equation}
 \alpha_{JKs} = \frac{(F_J/F_{Ks})}{(\nu_J/\nu_{Ks})},
\end{equation}
where $F_J$ and $F_{Ks}$ are fluxes calculated using the 2MASS zero values
from \citet{2003AJ....126.1090C} with respective central frequencies of these bands 
$\rm \nu_J$ and $\rm \nu_{Ks}$. The reddening corrections for the J, H, and Ks
bands are respectively 0.02149, 0.01332, and 0.00874 mag \citep[using $\rm R_V$ = 3.1 
and $\rm E(B - V) = 0.0241$]{1989ApJ...345..245C}. \\
\\
Figure 5 shows these spectral changes with time as well as with source flux states in the J band. 
Neither of these show any systematic trend over the long-term, as highlighted by the flat linear 
regression fits to them presented in Table 5. However, there are significant fluctuations around 
the mean, indicating spectral variations over short-time scales as reflected in the histograms 
shown in the right panels of figure \ref{fig:indx}. The histograms are skewed towards larger values 
of $\alpha_{Ks}$ indicating a tendency toward spectral steepening; however, there are a few instances 
showing spectral hardening. \\
\\
The fluxes vary over  almost an order of magnitude. The NIR SEDs, showing 
the diverse spectral facets exhibited by the source in between the minimum and
maximum NIR flux states are shown in Figure \ref{fig:video}. The accompanying video
presents a complete view of NIR SEDs with time. In general, the SEDs are flat
or declining, with most being consistent with a power-law spectrum (within a 10\%
error). Occasionally, there are hints of smooth departures at the low energy end as
well of hardening. 

\begin{deluxetable}{ccccc}
\tablenum{5}
\tablecaption{Spectral Index Variation with Respect to JD and J-band Magnitude for the Entire Period of the 
Observations of OJ 287 \label{tab:indexJKs}}
\tablewidth{0pt}
\tablehead{
\colhead{Parameter} & \colhead{$\rm m_2$} & \colhead{$\rm c_2$} & \colhead{$\rm r_2$} & \colhead{$\rm p_2$}
}
\startdata
$\alpha_{JKs}$ vs JD      & (-1.9$\rm\pm$1.4)$\times10^{-5}$  & 48$\rm\pm$33     & 0.006  & 0.16\\
$\alpha_{JKs}$ vs J (mag) & 0.014$\rm\pm$0.038                & 0.90$\rm\pm$0.49 & -0.006 & 0.72\\
\enddata

\vspace*{0.2cm}
\noindent
{\bf Note:} m$_{2}$ = slope and c$_{2}$ = intercept of $\alpha_{JKs}$ against JD or J; \\
r$_{2}$ = coefficient of determination ($\rm R^2$); p$_{2}$ (0.05) = null hypothesis rejection probability.
\end{deluxetable}

\subsubsection{Flaring and outbursts}
\noindent
During the nearly 14 years (2007 December -- 2021 November) of our intense multi-band NIR observations
of OJ 287 the source exhibited several well defined large amplitude flares seen  in all these 
J, H, and Ks bands, plotted in Figure 1 from bottom to top panels, respectively. We 
performed NIR inter-band cross correlation analysis using ZDCF and plotted this in Figure 2. 
From Figure 2, we found that J, H, and Ks bands fluxes are strongly correlated without 
any lag, so any observed flare in any of these J, H, and Ks bands are certainly observed 
quasi-simultaneously in the other two bands.    

\section{Discussion}\label{sec:discuss}

\noindent
The current study presents the most up-to-date and extensive NIR spectral and temporal 
behavior of OJ 287 for the lengthy period of 2007 December -- 2021 November.
Despite annual and inhomogeneous sampling related gaps, the NIR fluxes are well-sampled from 
high to low states, with denser sampling ($\sim 1-2$ days interval)
around and after the high states. This is true for almost every period of
activity, as is clear from Fig.\ \ref{fig:nirlc}. \\
\\
The source has undergone strong and quite frequent outbursts in NIR bands 
that are simultaneous within the observational cadence (Fig.\ \ref{fig:zdcf}). 
The respective magnitude histograms are skewed, with more gradual falloffs
on the brighter side but steeper declines on the fainter side. This skewness,
however, is most likely from a sampling bias favoring brighter state follow-up
and could also have a minor effect from the change of base level brightness,
as discussed in the next paragraph. The time series reveal strong NIR flux variations with
amplitudes almost similar to the optical bands of the same duration \citep{2012ApJ...756...13B,2014A&A...562A..79S,2017MNRAS.465.4423G,2019AJ....157...95G}. There is 
almost an order of magnitude difference between the extremes (see Fig.\ \ref{fig:video}). Over long-term 
timescales, there is no systematic spectral evolution or trend either with 
time (Fig.\ \ref{fig:tVsCor}) or flux state of the source (Fig.\ \ref{fig:HmagVsCor}). 
However, during the bright phases, the flux changes are often associated with significant 
color variations over the short-term, as highlighted by the fluctuations around the mean
in the color (Figs.\ \ref{fig:HmagVsCor} \& \ref{fig:tVsCor}) and spectral evolution plots 
(Figs.\ \ref{fig:indx} \& \ref{fig:video}). The color/spectral evolution with time and source 
brightness too are skewed, with a tendency for larger J-H color/spectral variations indicating 
steepening of the spectrum with source brightness over short-term flaring episodes. Contrary to 
this general trend, a few instances show appreciable hardening (Fig.\ \ref{fig:indx}).\\
\\
The behaviors reported here are largely in line with those reported previously for OJ 287
at NIR bands \citep[e.g.][]{1996A&AS..119..199Z,1998A&AS..133..163F,2012ApJ...756...13B,2014A&A...562A..79S}
and most of the seemingly contrary behavior can largely be
attributed to sampling bias of the previous studies and the change in base-level brightness. For example, the typical
brightness in J, H, and Ks bands are 12.9, 12.0, and 11.3 with a  typical standard
variation of $\sim 0.5$ mag in each and a $\rm 2.0 - 2.5$ mag difference between 
the extremes. These brightness levels
are in between the reported historical NIR brightness levels (1971 onwards) and 
so are the differences of the extremes \citep[$\sim 3.5$ mag;][]{1998A&AS..133..163F}.
However, since both the NIR and optical emissions are synchrotron and lie on the extension
of the same power-law spectral component (at and after the low-hump SED peak), the
century long optical light curve can be used to examine any systematic/trends. This
light curve indicates a systematic decline of base level brightness around 1 magnitude between 1971 and
2000 which reverses from 2000 onwards, with jet related short-term and large amplitude
flares superposed on it \citep[see Fig.\ 1 of][]{2018ApJ...866...11D}. 
Thus, the variations and differences between the extremes
are similar to those we see once the base brightness is taken into account. 
Similarly, the general tendency of larger J-K/J-Ks color (indicating steepening
of spectra) reported in earlier studies involving NIR and optical
data \citep[and references therein]{1996A&AS..119..199Z} is consistent with our
results during flaring. The long-term systematic trend reported in \citet{1996A&AS..119..199Z} is likely 
a sampling bias as is clear from the light curve which shows a systematic
decrease in flux before and after the most brightened event.\\
\\
The current NIR observations are also the first NIR data taken during the brightest
X-ray phases of this source that were seen in the years 2016--2017 and 2020 \citep{Komo20}
--- a result of a new high synchrotron peaked BL Lac (HBL) type of broadband emission component \citep{2018MNRAS.479.1672K,2021ApJ...921...18K,2022MNRAS.509.2696S}. Both these bright X-ray phases came after the claimed double-peaked
outbursts: the 2015 \citep{2016ApJ...819L..37V} and 2019 \citep{2020ApJ...894L...1L} flares of the $\rm\sim 12$-yr 
optical QPOs. As the  NIR variation amplitude is similar to that seen in the
optical \citep{2017MNRAS.465.4423G,2019AJ....157...95G} we can conclude that these
overall variations are due to a jet emission component rather than the new, thermal-like,
emission component seen during the 2013 -- 2016 at the interface of NIR-optical bands 
\citep{2018MNRAS.473.1145K}. This is also consistent with the brightest reported X-ray
phases of the source being an HBL-like emission component.\\
\\
Apart from these general trends, OJ 287 on short-terms at different activity
phases has shown very diverse and contrary behaviors. For example, none of the low
state SEDs presented here indicate any new emission component, but at most a spectral hardening;
however, on a few occasions, NIR-optical data show otherwise \citep{2014A&A...562A..79S}.
A hysterisis has also been reported involving 
redder-when-brighter and bluer-when-brighter trends as well as color changes at
fixed magnitude \citep{2012ApJ...756...13B}. The current observations
also make it clear that the extreme and odd variability seen
only in the K-band magnitude from the SMARTS\footnote{\url{www.astro.yale.edu/smarts/glast/home.php}} database 
that persisted for almost an observing cycle (JD: $\rm\sim$ 2455500 -- 2455710), as reported in \citet{2021arXiv211010851K}, 
is most likely artificial.
In short, although blazars are known for dynamic flux variability, they rarely
show significant spectral departures in the broadband SEDs. OJ 287, on the other hand,
is quite unique with sepctral changes persisting for much longer time \citep[e.g.][]
{2017ICRC...35..650B,2018MNRAS.473.1145K,2018MNRAS.479.1672K,2021ApJ...921...18K,
2021A&A...654A..38P,2022MNRAS.509.2696S} and thus, a potential source for fresh inputs not only on
relativistic jets above what is generally known about blazars
but also on aspects related to accretion as well \citep[e.g.][]{2020Galax...8...15K,
2021arXiv211010851K}.

\section{Summary}
\noindent
We have presented the most up-to date and extensive NIR observations of
OJ 287 between 2007 to 2021. A summary of our results and inferences
are as follows:

\begin{enumerate}
 \item OJ 287 shows strong NIR variations with a brightness changes
 of $\rm\gtrsim 2$ mag between the extremes. These variations are similar
 to those reported previously once the base level brightness is taken out, as 
 indicated by the optical light curve exceeding a century in length.
 
 \item The NIR variations are simultaneous within the limits of observational cadence.
 
 \item There is no general tendency for color variations over this extended period either with the flux 
 or with time. However, over short-times
 (bright phases) the NIR spectrum steepens with brightness and vice-versa.
 This tendency is similar to those reported in the literature in the 
 optical and NIR bands.
  
 \item A few of these observations show hardening of the NIR spectrum, possibly indicating
 a shift in the synchrotron SED peak, though they are not clearly significant. 
 
 \item The current NIR data includes the first data taken in these bands for bright X-ray phases.
  As those variabilities are similar to those in the optical they should arise from a broadband emission component.
\end{enumerate}

\section*{ACKNOWLEDGMENTS}

\noindent
The authors would like to dedicate this paper to the late Prof. S. S. Prasad who worked on exact solutions 
of Einstein's equations. Prof. S. S. Prasad is acknowledged for inspiring his son A.C.G, the first author of this paper. \\
\\
We thankfully acknowledge the anonymous reviewers for useful comments.
P.K.\ acknowledges support from the Department of Science and Technology (DST), government of India,
through the DST-INSPIRE Faculty grant (DST/INSPIRE/04/2020/002586). The INAOE, Mexico team thank CONACyT (Mexico) for
the research grant CB-A1-S-25070 (Y.D.M). H.G.X.\ is supported by the Ministry of Science 
and Technology of China (grant No.\ 2018YFA0404601) and the National Science Foundation of China 
(grants No.\ 11621303, 11835009, and 11973033). B.V. is funded by the Swedish Research Council (Vetenskapsr\aa det, 
grant no. 2017-06372). Z.Z.L. is thankful for support from the National 
Key R\&D Programme of China (grant No. 2018YFA0404602) and the Talented Program from the Chinese 
Academy of Sciences (CAS). \\
\\
{\it Facilities:} OAGH, CANICA. \\
\\
{\it Software:} {Astropy \citep{2013A&A...558A..33A,2018AJ....156..123A}, statsmodel \citep{2010Stats}, 
          DAOPHOT \citep{1987PASP...99..191S}, Gnuplot (version: 5.2; \url{http://www.gnuplot.info/}), IRAF \citep{1986SPIE..627..733T}
          }


\end{document}